\documentstyle[12pt]{article}

\begin{document}

\title{Curvature, Galactic Dynamics and Cosmic 
Repulsion\footnote{astro-ph/9811256, November 16, 1998}}

\author{\normalsize{Philip D. Mannheim} \\
\normalsize{Department of Physics,
University of Connecticut, Storrs, CT 06269} \\
\normalsize{mannheim@uconnvm.uconn.edu} \\}

\date{}

\maketitle

\begin{abstract}
We show that a cosmological negative spatial curvature can account for 
both a recently identified phenomenological imprint of the global Hubble 
flow on galactic rotation curves and for the recently detected cosmic 
repulsion and cosmic acceleration. Even though such a negative curvature
(a curvature which a galactic rest frame observer recognizes as a universal 
linear potential which gives rise to a universal acceleration) is only 
achievable in standard gravity by extreme fine tuning, nonetheless, it is 
naturally realizable in conformal gravity, a currently viable, fully 
covariant, candidate alternate gravitational theory. 
\end{abstract}

With the discovery of cosmic repulsion, the standard Newton-Einstein theory of
gravity now faces a quite severe new challenge, with the recent type Ia 
Supernovae high $z$ data requiring values for the parameters of the standard 
cosmology which are proving extremely difficult to justify. Specifically, 
the new data appear to have not only unambiguously excluded the highly popular 
inflationary universe paradigm of a $k=0$ spatially flat universe with a 
matter density $\Omega_{M}(t)$ which is always equal to one, but equally, once
$\Omega_{M}(t)$ is no longer equal to one, the only possible standard 
cosmologies which then are permitted by the data are all found to require some 
form or other of fine-tuning. Moreover, once $\Omega_{M}(t)$ is actually 
allowed to evolve in time, it would eventually have to depart radically from 
one in the very late universe, with its current closeness to one then being 
merely an artifact of the arbitrary time at which we actually are making 
observations. Hence its current era parameterization in terms of Newton's 
constant $G$ would appear to be accidental, to thus remove one of the primary 
reasons for believing that $G$ is necessarily a fundamental parameter which 
controls gravity on all scales and at all strengths. 

Given the seriousness of this situation we have suggested that the issue may 
not be that of trying to solve fine tuning problems at all, but rather that 
such problems may instead be an indicator that the standard cosmology might 
not in fact be the right one at all, with the extrapolation of standard 
gravity from its weak gravity solar system origins to altogether different 
distances and strengths simply being unreliable. Instead, we proposed an 
alternate, equally covariant, gravitational theory, viz. conformal gravity 
(a theory, interestingly, with no explicit fundamental $G$), a theory which 
departs from the standard theory precisely where the standard theory is having 
problems, to thus provide a very different extrapolation of standard 
Schwarzschild metric solar system wisdom. Characteristic of this alternate 
theory (see P. D. Mannheim, Phys. Rev. D58, 103511 (1998) for details) is that 
a $k<0$ topologically open universe is naturally favored (i.e. without fine 
tuning), with such a negatively curved universe then precisely producing the 
recently detected cosmic repulsion. Essentially, a space with negative 
curvature acts like a diverging refractive medium which causes particles 
to accelerate away from each other (positive curvature would lead to 
deceleration), with gravity (viz. curvature) itself then doing the repelling, 
with gravity thus acquiring an explicit repulsive component on cosmological 
distance scales, even as it still remains attractive on solar system ones.

Given such contrasting gravitational behaviors on cosmic and stellar distance 
scales, it is immediately natural to ask what happens at intermediate 
distances such as galactic, and, quite remarkably, it was found (P. D. 
Mannheim, ApJ 479, 659 (1997)) that then both effects are at play. 
Specifically, it was found that in 
conformal gravity stellar sources give rise to potentials of the form 
$V^{*}(r)=-\beta^{*}c^2/r+\gamma^{*}c^2r/2$, with the integration of both of 
these terms over the luminous matter distribution of a galaxy with
$N^{*}$ stars then producing a net acceleration $g_{loc}$ (viz. one which 
behaves asymptotically as $N^{*}\beta^{*}c^2/R^2+N^{*}\gamma^{*}c^2/2$) 
due to the local matter present in any given galaxy. At the same time, the 
global matter outside of each galaxy (viz. the Hubble flow) was then found to 
generate a further linear potential term, viz. the universal 
$g_{glob}=\gamma_{0}c^2/2$, coming precisely from the negative curvature of 
the universe according to $\gamma_{0}/2=(-k)^{1/2}$, so that the total 
acceleration on a test particle in any galaxy would then be given by 
$g_{tot}=g_{loc}+g_{glob}$, with the local, static, effect of global, 
comoving, cosmic repulsion thus being to push particles towards the centers of 
galaxies and cause their centripetal accelerations to increase. Further, 
through the use of this $g_{tot}$ parameter free (and dark matter free) 
fitting to a wide class of galactic rotation curves was then obtained  with an 
explicit value of $\gamma_{0}=3.06 \times 10^{-30}$ cm$^{-1}$ being extracted 
from the data. Galactic dynamics thus provides direct support for negative 
spatial curvature, with the inferred value for $(-k)^{1/2}$ encouragingly 
being found to precisely be a typical cosmologically significant length scale. 

Given this particular length scale we immediately see that the linear 
potential term first becomes competitive with the Newtonian one on none other
than galactic distance scales, to thus explain not only why solar system
physics is thus unaffected by conformal gravity, but also to identify at 
exactly what point departures from the luminous Newtonian contribution first 
set in. The spatial curvature of the universe thus fixes the scale at which
standard gravity needs to introduce dark matter in order to avoid failing to 
fit data. The emergence of the universal acceleration $\gamma_{0}c^2/2$ is
reminiscent of the acceleration $a_0$ of Milgrom's MOND theory.\footnote{The 
relation $a_0\simeq 4\gamma_{0}c^2\simeq 1.1\times 
10^{-8}$ cm s$^{-2}$ is found to hold numerically, with our relation 
$g_{tot}=g_{loc}+\gamma_{0}c^2/2$ even being a MOND type formula, though the 
one used in MOND fits is the somewhat different, ad hoc, non-analytic 
$g_{tot}=g_N\{1/2+(g^2_N+4a_0^2)^{1/2}/2g_N\}^{1/2}$ where $g_N$ is due to the 
Newtonian $V^{*}(r)=-\beta^{*}c^2/r$ potential alone.} Now while conformal 
gravity and MOND differ as to exactly how $g_{tot}$ is to depend on 
$\gamma_{0}$ or $a_0$ in the region where they are relevant, nonetheless it is 
important to note that both the theories recognize that it is precisely a 
universal acceleration with a cosmologically significant magnitude which is to 
set the scale at which departures from the luminous Newtonian expectation are 
to first set in (a fact which conformal gravity even 
derives).\footnote{Evidence for this selfsame 
acceleration has been found even in the solar system apparently, and if 
the Pioneer satellite data study of J. D. Anderson et al. Phys. Rev. 
Lett. 81, 2858 (1998) turns out to be correct, these data would then  
herald no less than a breakdown of Newtonian gravity within the solar 
system itself, so that no matter what one's views of alternate gravity, the 
case for extrapolating Newtonian gravity to larger distances without 
modification is severely weakened, as is then the very reason for believing in 
dark matter at all.} Since, as emphasized by S. S. McGaugh in these 
proceedings, galactic data certainly seem to be intimately aware of this onset 
scale, it would seem that if dark matter is to be correct, dark matter should 
then know about such a scale too, something not all that easy to achieve in a 
$k=0$ paradigm where there is no spatial curvature to potentially provide such 
a scale. Thus it would appear that a universal acceleration of magnitude 
$10^{-8}$ cm s$^{-2}$ is significant both cosmologically and galactically, and 
that the spatial curvature of the universe may not in fact be zero. This work 
has been supported in part by the Department of Energy under grant No. 
DE-FG02-92ER40716.00.

\end{document}